\documentclass[submission,Phys]{SciPost}
\usepackage{color,amsmath,amssymb,multirow,hyperref,booktabs,graphicx,mathtools} 

\renewenvironment{thebibliography}[1]{
  \begin{oldthebibliography}{#1}
    \setlength{\itemsep}{0em}
    \setlength{\parskip}{0em}
}
{
  \end{oldthebibliography}
}

\newcommand{\arxiv}[1]{\href{http://arxiv.org/abs/#1}{arXiv:#1}}

\newcommand\one{\leavevmode\hbox{\small1\normalsize\kern-.33em1}}

\newcommand{\gev}{\text{GeV}}

\def\slashchar#1{\setbox0=\hbox{$#1$}           
   \dimen0=\wd0                                 
   \setbox1=\hbox{/} \dimen1=\wd1               
   \ifdim\dimen0>\dimen1                        
      \rlap{\hbox to \dimen0{\hfil/\hfil}}      
      #1                                        
   \else                                        
      \rlap{\hbox to \dimen1{\hfil$#1$\hfil}}   
      /                                         
   \fi}

\newcommand{\eg}{\textsl{e.g.}\;}
\newcommand{\ie}{\textsl{i.e.}\;}

\newcommand{\be}{\begin{eqnarray*}}
\newcommand{\ee}{\end{eqnarray*}}

\newcommand{\bee}{\begin{eqnarray}}
\newcommand{\eee}{\end{eqnarray}}
\newcommand{\beeq}{\begin{equation}}
\newcommand{\eeeq}{\end{equation}}

\begin{document}

\begin{center}{\Large \textbf{
The Machine Learning Landscape of Top Taggers
}}\end{center}

\begin{center}
G.~Kasieczka (ed)\textsuperscript{1},
T.~Plehn (ed)\textsuperscript{2},
A.~Butter\textsuperscript{2},
K.~Cranmer\textsuperscript{3},
D.~Debnath\textsuperscript{4},
B.~M.~Dillon\textsuperscript{5},
M.~Fairbairn\textsuperscript{6},
D.~A.~Faroughy\textsuperscript{5},
W.~Fedorko\textsuperscript{7},
C.~Gay\textsuperscript{7},
L.~Gouskos\textsuperscript{8},
J.~F.~Kamenik\textsuperscript{5,9},
P.~T.~Komiske\textsuperscript{10},
S.~Leiss\textsuperscript{1},
A.~Lister\textsuperscript{7},
S.~Macaluso\textsuperscript{3,4},
E.~M.~Metodiev\textsuperscript{10},
L.~Moore\textsuperscript{11},
B.~Nachman,\textsuperscript{12,13},
K.~Nordstr\"om\textsuperscript{14,15},
J.~Pearkes\textsuperscript{7},
H.~Qu\textsuperscript{8},
Y.~Rath\textsuperscript{16},
M.~Rieger\textsuperscript{16},
D.~Shih\textsuperscript{4},
J.~M.~Thompson\textsuperscript{2}, and
S.~Varma\textsuperscript{6}
\end{center}

\begin{center}
{\bf 1} Institut f\"ur Experimentalphysik, Universit\"at Hamburg, Germany\\
{\bf 2} Institut f\"ur Theoretische Physik, Universit\"at Heidelberg, Germany \\
{\bf 3} Center for Cosmology and Particle Physics and Center for Data Science, NYU, USA \\
{\bf 4} NHECT, Dept. of Physics and Astronomy, Rutgers, The State University of NJ, USA \\
{\bf 5} Jozef  Stefan  Institute, Ljubljana,  Slovenia  \\
{\bf 6} Theoretical Particle Physics and Cosmology, King’s College London, United Kingdom \\
{\bf 7} Department of Physics and Astronomy, The University of British Columbia, Canada \\
{\bf 8} Department of Physics, University of California, Santa Barbara, USA \\
{\bf 9} Faculty of Mathematics and Physics, University  of  Ljubljana, Ljubljana, Slovenia \\
{\bf 10} Center for Theoretical Physics, MIT, Cambridge, USA \\
{\bf 11} CP3, Universit{\'exx} Catholique de Louvain, Louvain-la-Neuve, Belgium\\
{\bf 12} Physics Division, Lawrence Berkeley National Laboratory, Berkeley, USA \\
{\bf 13} Simons Inst. for the Theory of Computing, University of California, Berkeley, USA\\
{\bf 14} National Institute for Subatomic Physics (NIKHEF), Amsterdam, Netherlands \\
{\bf 15} LPTHE, CNRS \& Sorbonne Universit\'e, Paris, France \\
{\bf 16 } III. Physics Institute A, RWTH Aachen University, Germany \\[3mm]

gregor.kasieczka@uni-hamburg.de \\
plehn@uni-heidelberg.de
\end{center}

\begin{center}
\today
\end{center}

\section*{Abstract}
{\bf Based on the established task of identifying boosted,
  hadronically decaying top quarks, we compare a wide range of modern
  machine learning approaches. Unlike most established methods they
  rely on low-level input, for instance calorimeter output. While
  their network architectures are vastly different, their performance
  is comparatively similar. In general, we find that these new
  approaches are extremely powerful and great fun.}

\clearpage

\vspace{10pt}
\noindent\rule{\textwidth}{1pt}
\tableofcontents\thispagestyle{fancy}
\noindent\rule{\textwidth}{1pt}
\vspace{10pt}

\clearpage

\section{Introduction}
\label{sec:intro}

Top quarks are, from a theoretical perspective, especially interesting
because of their strong interaction with the Higgs boson and the
corresponding structure of the renormalization group. Experimentally,
they are unique in that they are the only quarks which decay before
they hadronize. One of the qualitatively new aspect of LHC physics are
the many signal processes which for the first time include phase space
regimes with strongly boosted tops. Those are typically analyzed with
the help of jet algorithms~\cite{Seymour:1993mx}. Corresponding jet
substructure analyses have found their way into many LHC measurements
and searches.

Top tagging based on an extension of standard jet algorithms has a
long history~\cite{Butterworth:2008iy,toptags_history}.  Standard top
taggers used by ATLAS and CMS usually search for kinematic
features induced by the top and $W$-boson
masses~\cite{Kaplan:2008ie,Plehn:2009rk,Plehn:2010st}. This implies
that top tagging is relatively straightforward, can be described in
terms of perturbative quantum field theory, and hence makes an obvious
candidate for a benchmark process. An alternative way to tag tops is
based on the number of prongs, properly defined as
$N$-subjettiness~\cite{Thaler:2010tr}. Combined with the SoftDrop mass
variable~\cite{Larkoski:2014wba}, this defines a particularly economic
2-parameter tagger, but without a guarantee that the full top momentum
gets reconstructed.

Based on simple deterministic taggers, the LHC collaborations have
established that subjet analyses work and can be controlled in their
systematic uncertainties~\cite{boost}.  The natural next step are
advanced statistical methods~\cite{shower_deco}, including
multi-variate analyses~\cite{Kasieczka:2015jma}. In the same spirit,
the natural next question is why we apply highly complex
tagging algorithms to a pre-processed set of kinematic observables
rather than to actual data. This question becomes especially relevant
when we consider the significant conceptual and performance progress
in machine learning.  Deep learning, or the use of neural networks
with many hidden layers, is the tool which allows us to analyze
low-level LHC data without constructing high-level observables. This
directly leads us to standard classification tools in contemporary
machine learning, for example in image or language
recognition.

The goal of this study is to see how well different neutral network
setups can classify jets based on calorimeter information. A
straightforward way to apply standard machine learning tools to jets
is so-called calorimeter images, which we use for our comparison of
the different available approaches on an equal footing. Considering
calorimeter cells inside a fat jet as pixels defines a sparsely filled
image which can be analyzed through standard convolutional
networks~\cite{Cogan:2014oua,deOliveira:2015xxd,Baldi:2016fql}. A set
of top taggers defined on the basis of image recognition will be part
of our study and will be described in
Sec.~\ref{sec:methods_image}~\cite{Kasieczka:2017nvn,Macaluso:2018tck}. A
second set of taggers is based directly on the 4-momenta of the subjet
constituents and will be introduced in
Sec.~\ref{sec:methods_vec}~\cite{Almeida:2015jua,Pearkes:2017hku,Erdmann:2018shi};
recurrent neural networks inspired by language
recognition~\cite{Louppe:2017ipp} can be grouped into the same
category.  Finally, there are taggers which are motivated by
theoretical considerations like soft and collinear radiation patterns
or infrared
safety~\cite{Komiske:2017aww,Butter:2017cot,Komiske:2018cqr,Moore:2018lsr}
which we collect in Sec.~\ref{sec:methods_th}.\medskip

While initially it was not clear if any of the machine learning
methods applied to top tagging would be able to significantly exceed
the performance of the multi-variate
tools~\cite{Almeida:2015jua,Kasieczka:2017nvn}, later studies have
consistently showed that we can expect great performance
improvement from most modern tools. This turns around the question
into which of the tagging approaches have the best performance (also
relative to their training effort), and if the leading taggers make
use of the same, hence complete set of information. Indeed, we will
see that we can consider jet classification based on deep learning at
the pure performance level an essentially solved problem. For a
systematic experimental application of these tools our focus will be
on a new set of questions related to training data, benchmarking,
calibration, systematics, etc.

\section{Data set}
\label{sec:sample}

The top signal and mixed quark-gluon background jets are produced with
using Pythia8~\cite{Sjostrand:2014zea} with its default tune for a
center-of-mass energy of 14 TeV and ignoring multiple interactions and
pile-up.  For a simplified detector simulation we use
Delphes~\cite{deFavereau:2013fsa} with the default ATLAS detector
card. This accounts for the curved trajectory of the charged
particles, assuming a magnetic field of 2~T and a radius of 1.15~m as
well as how the tracking efficiency and momentum smearing changes with
$\eta$. The fat jet is then defined through the anti-$k_T$
algorithm~\cite{Cacciari:2008gp} in FastJet~\cite{Cacciari:2011ma}
with $R = 0.8$. We only consider the leading jet in each event and
require
\begin{align}
 p_{T,j} = 550~....~650~\gev \; .
\end{align}
For the signal only, we further require a matched parton-level top to
be within $\Delta R = 0.8$, and all top decay partons to be within
$\Delta R = 0.8$ of the jet axis as well. No matching is performed for
the QCD jets. We also require the jet to have $|\eta_j| < 2$. The
constituents are extracted through the Delphes energy-flow algorithm,
and the 4-momenta of the leading 200 constituents are stored. For jets
with less than 200 constituents we simply add zero-vectors.


\begin{figure}[b!]
  \includegraphics[width=0.325 \textwidth]{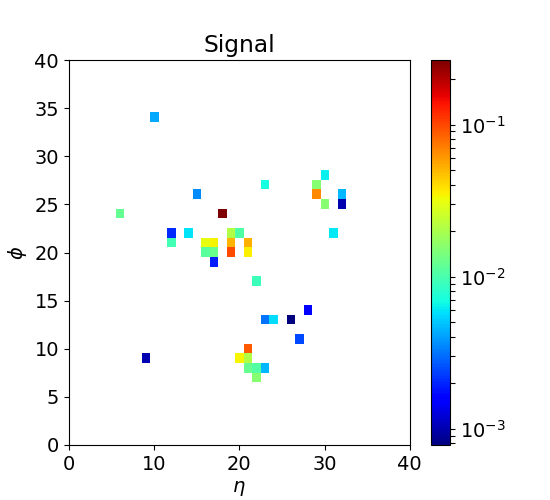}
  \includegraphics[width=0.325 \textwidth]{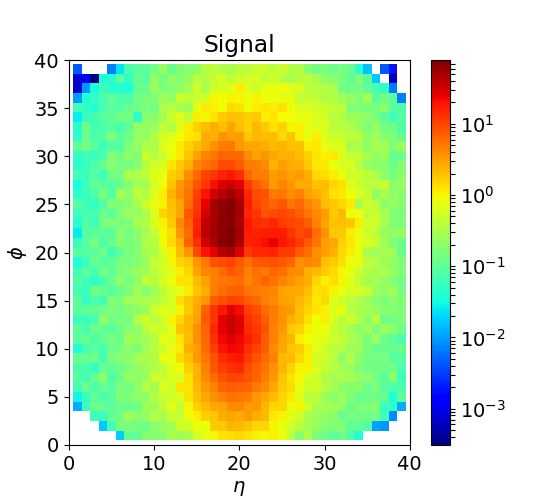}
  \includegraphics[width=0.325 \textwidth]{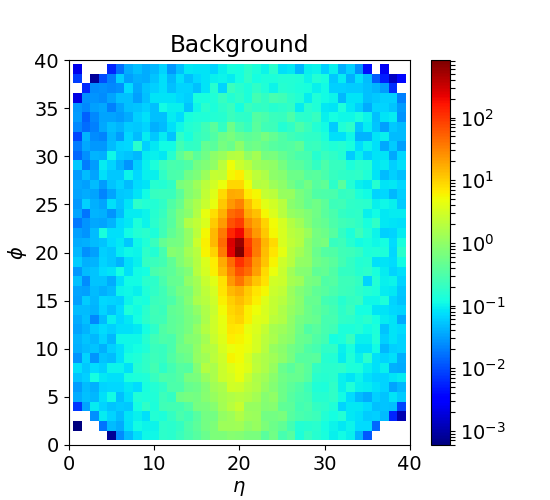}
  \caption{Left: typical single jet image in the rapidity vs azimuthal
    angle plane for the top signal after pre-processing. Center and
    right: signal and background images averaged over 10,000
    individual images.}
  \label{fig:averaged_images}
\end{figure} 

Particle information or additional tracking information is not
included in this format. For instance, we do not record charge
information or the expected displaced vertex from the
$b$-decay. Therefore, the quoted performance should not be considered
the last word for the LHC. On the other hand, limiting ourselves to
essentially calorimeter information allows us to compare many
different techniques and tools on an equal footing.\medskip


Our public data set consists of 1~million signal and 1~million
background jets and can be obtained from the authors upon
request~\cite{dataset}. They are divided into three samples: training with 600k
signal and background jets each, validation with 200k signal and
background jets each, and testing with 200k signal and 200k background
jets. For proper comparison, all algorithms are optimized using the
training and validation samples and all results reported are obtained
using the test sample. For each algorithm, the classification result
for each jet is made available, so we can not only measure the
performance of the network, but also test which jets are correctly
classified in each approach.

\section{Taggers}
\label{sec:methods}

\subsection{Imaged-based taggers}
\label{sec:methods_image}

To evaluate calorimeter information efficiently we can use powerful
methods from image recognition. We simply interpret the energy
deposition in the pixelled calorimeter over the area of the fat jet as
an image and apply convolutional networks to it. These convolutional
networks encode the 2-dimensional information which the network would
have to learn if we gave it the energy deposition as a 1-dimensional
chain of entries. Such 2-dimensional networks are the drivers behind
many advances in image recognition outside physics and allow us to
benefit from active research in the machine learning community.

If we approximate the calorimeter resolution as $0.04 \times
2.25^\circ$ in rapidity vs azimuthal angle a fat jet with radius
parameter $R=0.8$ can be covered with $40 \times 40$ pixels. Assuming
a $p_T$ threshold around 1~GeV, a typical QCD jet will feature around
40 constituents in this jet image~\cite{Butter:2017cot}.  In
Fig.~\ref{fig:averaged_images} we show an individual calorimeter image
from a top jet, as well as averaged images of top jets and QCD jets,
after some pre-processing. For both, signal and background jets the
center of the image is defined by the hardest object. Next, we rotate
the second-hardest object to 12 o'clock. Combined with a narrow
$p_T$-bin for the jets this second jet develops a preferred distance
from the center for the signal but not for the QCD
background. Finally, we reflect the third-largest object to the right
side of the image, where such a structure is really only visible for
the 3-prong top signal. Note that this kind of pre-processing is
crucial for visualization, but not necessarily part of a tagger. We
will compare two modern deep networks analyzing calorimeter images. A
comparison between the established multi-variate taggers and the
modestly performing early DeepTop network can be found in
Ref.~\cite{Kasieczka:2017nvn}.

\subsubsection{CNN}

\begin{figure}[b!]
  \includegraphics[width=1.0\textwidth]{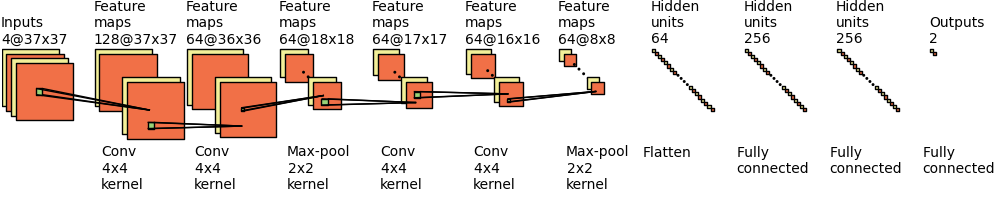}
\caption{Architecture of the CNN top tagger. Figure from
  Ref.~\cite{Macaluso:2018tck}.}
\label{fig:NNarch}
\end{figure}

One standard top tagging method applies a convolutional neural network
(CNN) trained on jet images, generated from the list of per-jet
constituents of the reference sample~\cite{Macaluso:2018tck}. We
perform a specific preprocessing before pixelating the image.  First,
we center and rotate the jet according to its $p_T$-weighted centroid
and principal axis. Then we flip horizontally and vertically so that
the maximum intensity is in the upper right quadrant. Finally, we
pixelate the image with $p_T$ as the pixel intensity, and normalize it
to unit total intensity. Although our original method includes color
images, where the track $p_T$'s and neutral $p_T$'s are considered
separately, for this dataset we restrict ourselves to gray-scale
images.

For the design of our CNN, we take the DeepTop
architecture~\cite{Kasieczka:2017nvn} as a starting point, but augment
it with more feature maps, hidden units, etc. The complete network
architecture is illustrated in Fig.~\ref{fig:NNarch}.

The CNN is implemented on an NVidia Tesla P100 GPU using PyTorch.  For
training we use the cross entropy loss function and Adam as the
optimizer~\cite{Kingma:2014vow} with a minibatch size of 128.  The
initial learning rate is $5\times 10^{-5}$ and is decreased by a
factor of 2 every 10 epochs. The CNN is trained for 50 epochs and the
epoch with the best validation accuracy is chosen as the final tagger.

\subsubsection{ResNeXt}

The ResNeXt model is another deep CNN using jet images as
inputs. The images used in this model are 64$\times$64 pixels in size
centered on the jet axis, corresponding to a granularity of 0.025
radians in the $\eta-\phi$ space. The intensity of each pixel is the
sum of $p_T$ of all the constituents within the pixel. The CNN
architecture is based on the 50-layer ResNeXt
architecture~\cite{DBLP:journals/corr/XieGDTH16}.  To adapt to the
smaller size of the jet images, the number of channels in all the
convolutional layers except for the first one is reduced by a factor
of 4, and a dropout layer with a keep probability of 0.5 is added
after the global pooling. The network in implemented in Apache
MXNet~\cite{DBLP:journals/corr/ChenLLLWWXXZZ15}, and trained from
scratch on the top tagging dataset.

\subsection{4-Vector-based taggers}
\label{sec:methods_vec}

A problem with the image recognition approaches discussed in
Sec.~\ref{sec:methods_image} arises when we want to include additional
information for example from tracking or particle identification. We
can always combine different images in one
analysis~\cite{Gallicchio:2010sw}, but the significantly different
resolution for example of calorimeter and tracker images becomes a
serious challenge. For a way out we can follow the experimental
approach developed by CMS and ATLAS and use particle flow or similar
objects as input to neural network taggers. In our case this means
4-vectors with the energy and the momentum of the jet
constituents. The challenge is to define an efficient network setup
that either knows or learns the symmetry properties of 4-vectors and
replace the notion of 2-dimensional geometric structure included in
the convolutional networks

Of the known, leading properties of top jets the 4-vector approach can
be used to efficiently extract the number of prongs as well as the
number of constituents as a whole. For the latter, it is crucial for
the taggers to notice that soft QCD activity is universal and should
not provide additional tagging information.

\begin{figure}[b!]
\includegraphics[width=0.45\textwidth]{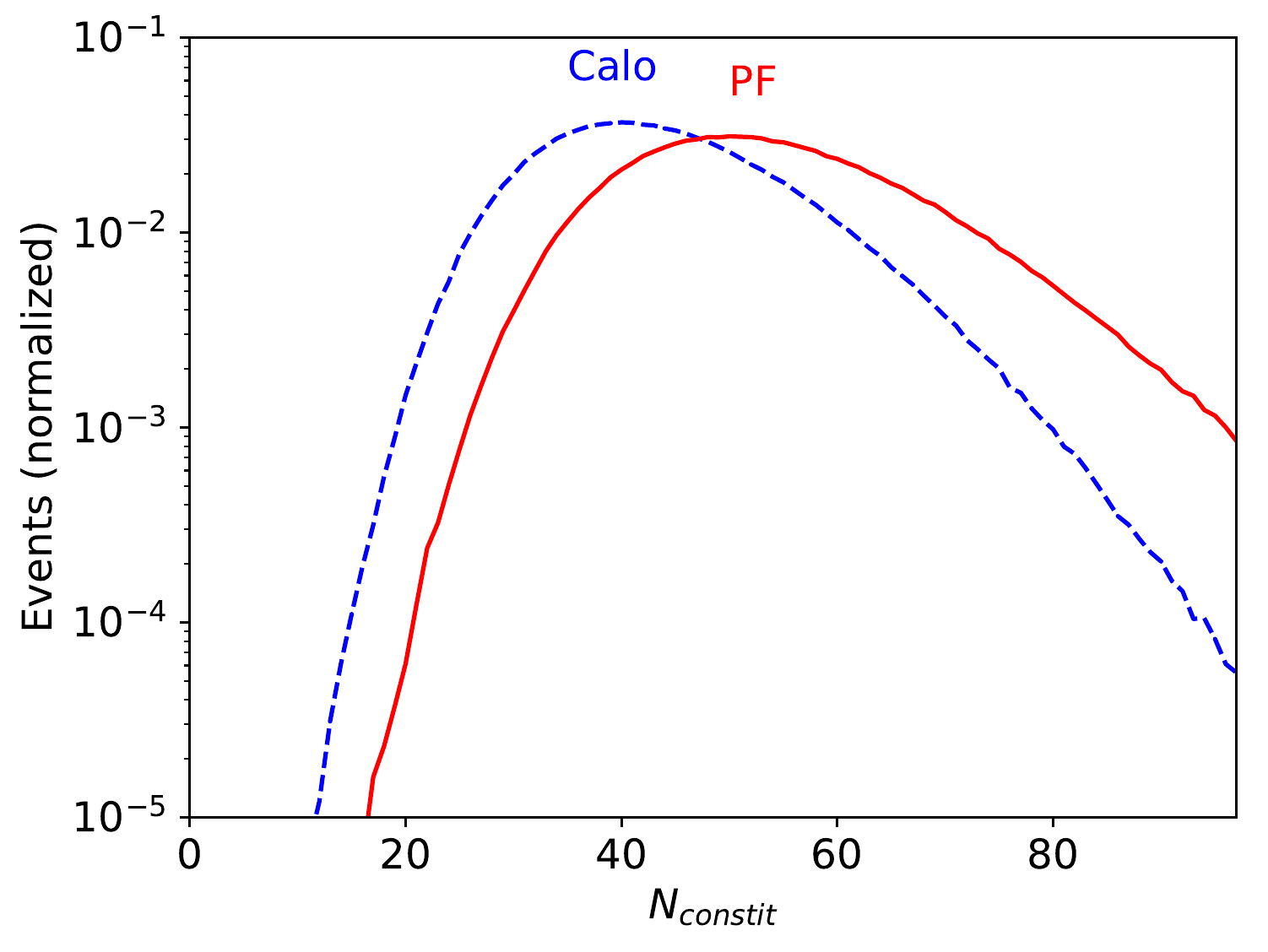}
\hspace*{0.1\textwidth}
\includegraphics[width=0.45\textwidth]{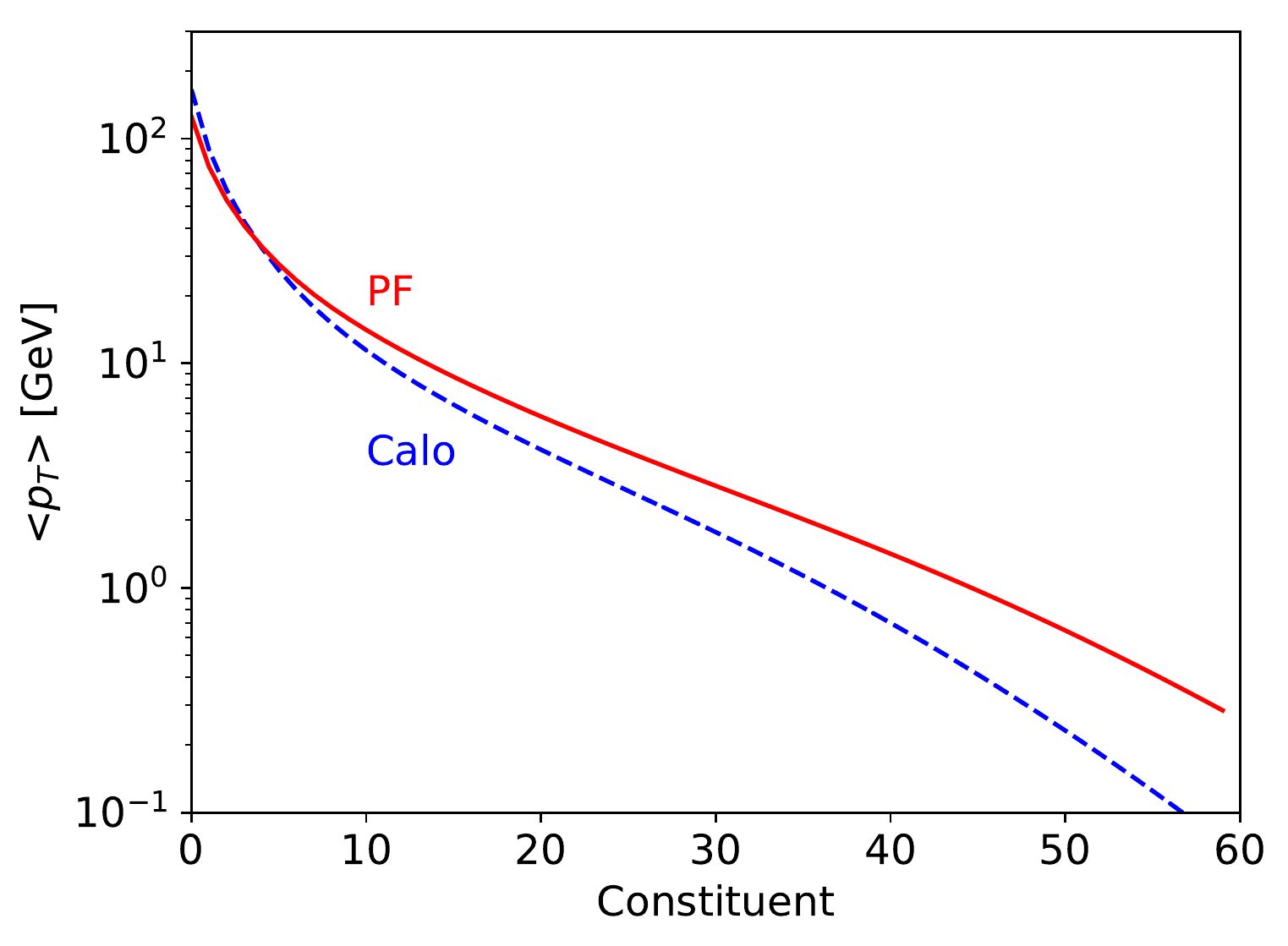}
\caption{Number of constituents (left) and mean of the transverse
  momentum (right) of the ranked constituents of a typical top jet. We
  show calorimeter entries as well as particle flow constituents after
  Delphes.}
\label{fig:nconst}
\end{figure}

\subsubsection{TopoDNN}

If we start with 200 $p_T$-sorted 4-vectors per jet, the arguably
simplest deep network architecture is a dense network taking all 800
floating point numbers as a fixed set~\cite{Pearkes:2017hku}.  Since
the mass of individual particle-flow candidates cannot be reliably
reconstructed, the TopoDNN tagger uses at most 600 inputs, namely
$(p_T,\eta,\phi)$ for each constituent. To improve the training
through physics-motivated pre-processing, a longitudinal boost and a
rotation in the transverse plane are applied such that the $\eta$ and
$\phi$ values of the highest-$p_T$ constituent is centered at $(0,0)$.
The momenta are scaled by a constant factor 1/1700, chosen ad-hoc
because a dynamic quantity such as the constituent $p_T$ can distort
the jet mass~\cite{deOliveira:2015xxd}.  A further rotation is applied
so that the second highest jet constituent is aligned with the
negative $y$-axis, to remove the rotational symmetry of the second
prong about the first prong in the jet.  This is a proper rotation,
not a simple rotation in the $\eta$-$\phi$ plane, and thus preserves
the jet mass (but can distort quantities like
$N$-subjettiness~\cite{Thaler:2010tr,deOliveira:2017pjk}).

The TopoDNN tagger presented here has a similar architecture as a
tagger with the same name used by the ATLAS
collaboration~\cite{Aaboud:2018psm}. The architecture of this TopoDNN
tagger was optimized for a dataset of high $p_T$ (450 to 2400~GeV)
$R=1.0$ trimmed jets.  Its hyper-parameters are the number of
constituents considered as inputs, the number of hidden layers, the
number of nodes per layer, and the activation functions for each
layer.  For the top dataset of this study we find that 30 constituents
saturate the network performance.  The ATLAS tagger uses only 10 jet
constituents, but the inputs are topoclusters~\cite{Aad:2016upy} and
not individual particles. The remaining ATLAS TopoDNN architecture is
not altered.

Individual particle-flow candidates in the experiment are also not individual
particles, but there is a closer correspondence. For this reason, the
TopoDNN tagger performance presented here is not directly comparable
to the results presented in Ref.~\cite{Aaboud:2018psm}.

\subsubsection{Multi-Body N-Subjettiness}

The multi-body phase space tagger is based on the proposal in
Ref.~\cite{Datta:2017rhs} to use a basis of $N$-subjettiness
variables~\cite{Thaler:2010tr} spanning an $m$-body phase space to
teach a dense neural network to separate background and signal using a
minimal set of input variables.  A setup for the specific purpose of
top tagging was introduced in Ref.~\cite{Moore:2018lsr}. To generate
the input for our network we first convert the event to HEPMC and use
Rivet~\cite{Buckley:2010ar} to evaluate $N$-subjettiness variables by
using the FastJet~\cite{Cacciari:2011ma} contrib plug-in
library~\cite{Thaler:2010tr,Thaler:2011gf}. The input variables
spanning the $m$-body phase space are then given by:
\begin{equation}
\left\{ \tau_1^{(0.5)},\tau_1^{(1)},\tau_1^{(2)},\tau_2^{(0.5)},\tau_2^{(1)},\tau_2^{(2)}, \dots, \tau_{m-2}^{(0.5)},\tau_{m-2}^{(1)},\tau_{m-2}^{(2)},\tau_{m-1}^{(1)},\tau_{m-1}^{(2)}  \right\}, \label{eq:mbodyinput}
\end{equation}
where
\begin{equation}
\tau_N^{(\beta)} = \frac{1}{p_{T,J}} \sum_{i \in J} p_{T,i} \min \left\{R_{1i}^\beta,R_{2i}^\beta, \dots, R_{Ni}^\beta \right\},
\label{eq:nsubdef}
\end{equation}
and $R_{ni}$ is the distance in the $\eta - \phi$ plane of the jet
constituent $i$ to the axis $n$. We choose the $N$ jet axes using the
$k_T$ algorithm~\cite{Ellis:1993tq} with $E$-scheme recombination,
specifically $N=6,8$ for our explicit comparison. This input has the
advantage that it is theoretically sound and IR-safe, while it can be
easily understood as a set of correlators between 4-momenta combining
the number of prongs in the jet with additional momentum information.

The machine learning setup is a dense neural network implemented in
TensorFlow~\cite{tensorflow} using these input variables. We also add
the jet mass and jet $p_T$ as input variables to allow the network to
learn physical scales. The network consists of four fully connected
hidden layers, the first two with 200 nodes and a dropout
regularization of 0.2, and the last two with 50 nodes and a dropout
regularization of 0.1. The output layer consists of two nodes. We use
a ReLu activation function throughout and minimize the cross-entropy
using Adam optimization~\cite{Kingma:2014vow}.

\subsubsection{TreeNiN}

In this method, a tree neural network (TreeNN) is trained on jet
trees. The TreeNN provides a jet embedding, which maps a set of
4-momenta into a vector of fixed size and can be trained together with
a successive network used for classification or
regression~\cite{Louppe:2017ipp}.  Jet constituents of the reference
sample are reclustered to form binary trees, and the topology is
determined by the clustering algorithm, \ie $k_T$, anti-$k_T$ or
Cambridge/Aachen.  For this paper, we choose the $k_T$ clustering
algorithm, and 7 features for the nodes: $|p|$, $\eta$, $\phi$, $E$,
$E/E_{\text{jet}}$, $p_T$ and $\theta$.  We scale each feature with
the Scikit-learn preprocessing method RobustScaler, which is
robust to outliers~\cite{scikit-learn}.

To speed up the training, a special batching is implemented in
Ref.~\cite{Louppe:2017ipp}. Jets are reorganized by levels, \eg the root
node of each tree in the batch is added at level zero, their children
at level one, etc. Each level is restructured such that all internal
nodes come first, followed by outer nodes (leaves), and zero padding
is applied when necessary. We developed a PyTorch implementation to
provide GPU acceleration.
 
We introduce in Ref.~\cite{TreeNiNpaper} a Network-in-Network
generalization of the simple TreeNN architecture proposed in
Ref.~\cite{Louppe:2017ipp}, where we add fully connected layers at
each node of the binary tree before moving forward to the next
level. We refer to this model as TreeNiN. In particular, we add 2 NiN
layers with ReLU activations. Also, we split weights between internal
nodes and leaves, both for the NiN layers and for the initial
embedding of the 7 input features of each node. Finally, we introduce
two sets of independent weights for the NiN layers of the left and
right children of the root node.  The code is publicly accessible on
GitHub~\cite{RecNN}.  Training is performed over 40 epochs with a
minibatch of 128 and a learning rate of $2\,$x$\,10^{-3}$ decayed by a
factor of 0.9 after every epoch, using the cross entropy loss function
and the Adam optimizer~\cite{Kingma:2014vow}.

\subsubsection{P-CNN}

The particle-level convolutional neural network (P-CNN), used in the
CMS particle-based DNN tagger~\cite{CMS-DP-2017-049}, is a customized
1-dimensional CNN for boosted jet tagging. Each input jet is
represented as a sequence of constituents with a fixed length of 100,
organized in descending order of $p_T$. The sequence is padded with
zeros if a jet has less than 100 constituents. If a jet contains more
than 100 constituents, the extra constituents are discarded. For each
constituent, seven input features are computed from the 4-momenta of
the constituent and used as inputs to the network: $\log p_T^i$, $\log
E^i$, $\log({p_T^i}/{p_T^\text{jet}})$, $\log(E^i/E^\text{jet})$,
$\Delta\eta^i$, $\Delta\phi^i$ and $\Delta R^i$. Angular distances are
measured with respect to the jet axis. The use of these transformed
features instead of the raw 4-momenta was found to lead to slightly
improved performance.

The P-CNN used in this paper follows the same architecture as the CMS
particle-based DNN tagger. However, the top tagging dataset in this
paper contains only kinematic information of the particles, while the
CMS particle-based DNN tagger uses particle tracks and secondary
vertices in addition. Therefore, the network components related to
them are removed for this study. The P-CNN is similar to the ResNet
model~\cite{DBLP:journals/corr/HeZR016} for image recognition, but
only uses a 1-dimensional convolution instead of 2-dimensional
convolutions. One of the features that distinguishes the ResNet
architecture from other CNNs is that it includes skip connections
between layers. The number of convolutional layers is 14, all with a
kernel size of 3. The number of channels for the 1-dimensional
convolutional layers ranges from 32 to 128. The outputs of the
convolutional layers undergo a global pooling, followed by a
fully-connected layer of 512 units and a dropout layer with a keep
rate of 0.5, before yielding the final prediction. The network is
implemented in Apache MXNet~\cite{DBLP:journals/corr/ChenLLLWWXXZZ15}
and trained with the Adam optimizer~\cite{Kingma:2014vow}.

\subsubsection{ParticleNet}

Similar to the Particle Flow Network, the
ParticleNet~\cite{Qu:2019gqs} is also built on the point cloud
representation of jets, where each jet is treated as an unordered set
of constituents. The input features for each constituent are the same
as that in the P-CNN model. The ParticleNet first constructs a graph
for each jet, with the constituents as the vertices. The edges of the
graph are then initialized by connecting each constituent to its $k$
nearest-neighbor constituents based in $\eta-\phi$ space. The EdgeConv
operation~\cite{DBLP:journals/corr/abs-1801-07829} is then applied on
the graph to transform and aggregate information from the nearby
constituents at each vertex, analogous to how regular convolution
operates on square patches of images. The weights of the EdgeConv
operator are shared among all constituents in the graph, therefore
preserving the permutation invariance property of the constituents in a
jet. The EdgeConv operations can be stacked to form a deep graph
convolutional network.

The ParticleNet relies on the dynamic graph convolution approach of
Ref.~\cite{DBLP:journals/corr/abs-1801-07829} and further extends
it. The ParticleNet consists of three stages of EdgeConv operations,
with three EdgeConv layers and a shortcut
connection~\cite{DBLP:journals/corr/HeZRS15} at each stage. Between
the stages, the jet graph is dynamically updated by redefining the
edges based on the distances in the new feature space generated by the
EdgeConv operations. The number of nearest neighbors, $k$, is also
varied in each stage. The three stages of EdgeConv are followed by a
global pooling over all constituents, and then two fully connected
layers. The details of the ParticleNet architecture can be found
in~\cite{Qu:2019gqs}. It is implemented in Apache
MXNet~\cite{DBLP:journals/corr/ChenLLLWWXXZZ15} and trained with
Adam~\cite{Kingma:2014vow}.

\subsection{Theory-inspired taggers}
\label{sec:methods_th}

Going beyond a relatively straightforward analysis of 4-vectors we can
build networks specifically for subjet analyses and include as much of
our physics knowledge as possible. The motivation for this is
two-fold: building this information into the network should save
training time, and it should allow us to test what kind of physics
information the network relies on.

At the level of these 4-vectors the main difference between top jets
and massless QCD jets is two mass
drops~\cite{Butterworth:2008iy,Plehn:2009rk}, which appear after we
combine 4-vectors based on soft and collinear proximity. While it is
possible for taggers to learn Lorentz boosts and the Minkowski metric,
it might be more efficient to give this information as part of the
tagger input and architectures.

In addition, any jet analysis tool should give stable results in the
presence of additional soft or collinear splittings. From theory we
know that smooth limits from very soft or collinear splittings to no
splitting have to exist, a property usually referred to as infrared
safety. If we replace the relatively large pixels of calorimeter
images with particle-level observables it is not clear how IR-safe a
top tagging output really is~\cite{Choi:2018dag}. This is not only a
theoretical problem which arises when we for example want to compare
rate measurements with QCD predictions, a lack of IR-safety will also
make it hard to train or benchmark taggers on Monte Carlo simulations
and to extraction tagging efficiencies using Monte Carlo
input~\cite{Barnard:2016qma}.

\subsubsection{Lorentz Boost Network}

The Lorentz Boost Network (LBN) is designed to autonomously extract a
comprehensive set of physics-motivated features given only low-level
variables in the form of constituent 4-vectors~\cite{Erdmann:2018shi}.
These engineered features can be utilized in a subsequent neural
network to solve a specific physics task.  The resulting two-stage
architecture is trained jointly so that extracted feature
characteristics are adjusted during training to serve the minimization
of the objective function by means of back-propagation.

The general approach of the LBN is to reconstruct parent particles
from 4-vectors of their decay products and to exploit their properties
in appropriate rest frames.  Its architecture is comprised of three
layers.  First, the input vectors are combined into $2 \times M$
intermediate vectors through linear combinations.  Corresponding
linear coefficients are trainable and constrained to positive numbers
to prevent constructing vectors with unphysical implications, such as
$E < 0$ or $E < m$.  In the subsequent layer, half of these
intermediate vectors are treated as constituents, whereas the other half
are considered rest frames.  Via Lorentz transformation the constituents
are boosted into the rest frames in a pairwise approach, \ie the
$m^\text{th}$ constituent is boosted into the $m^\text{th}$ rest frame.
In the last layer, $F$ features are extracted from the obtained $M$
boosted constituents by employing a set of generic feature mappings.  The
autonomy of the LBN lies in its freedom to construct arbitrary boosted
particles through trainable particle and rest frame combinations, and
to consequently access and provide underlying characteristics which
are otherwise distorted by relativistic kinematics.

The order of input vectors is adjusted before being fed to the LBN.
The method utilizes linearized clustering histories as preferred by
the anti-$k_T$ algorithm~\cite{Cacciari:2008gp}, implemented in
Fastjet~\cite{Cacciari:2011ma}.  First, the jet constituents are
reclustered with $\Delta R = 0.2$, and the resulting subjets are
ordered by $p_T$.  Per subjet, another clustering with $\Delta R =
0.2$ is performed and the order of constituents is inferred from the
anti-$k_T$ clustering history.  In combination, this approach yields a
consistent order of the initial jet constituents.

The best training results for this study are obtained for $M = 50$ and
six generic feature mappings: $E$, $m$, $p_T$, $\phi$, and $\eta$ of
all boosted constituents, and the cosine of the angle between momentum
vectors of all pairs of boosted constituents, in total $F = 5 M + (M^2
- M) / 2$ features.  Batch normalization with floating averages during
training is employed after the feature extraction
layer~\cite{Ioffe:2015ovl}.  This subsequent neural network
incorporates four hidden layers, involving 1024, 512, 256, and 128
exponential linear (ELU) units, respectively.  Generalization and
overtraining suppression are enforced via L2 regularization with a
factor of $10^{-4}$.  The Adam optimizer is utilized for minimizing
the binary cross-entropy loss~\cite{Kingma:2014vow}, configured with
an initial learning rate of $10^{-3}$ for a batch size of 1024.  The
training procedure converges after around 5000 batch iterations.

\subsubsection{Lorentz Layer}

Switching from image recognition to a setup based on 4-momenta we can
take inspiration from graph convolutional networks. Also used for the
ParticleNet they allow us to analyze sparsely filled images in terms
of objects and a free distance measure. While usually the most
appropriate distance measure needs to be determined from data,
fundamental physics tells us that the relevant distance for
jet physics is the scalar product of two 4-vectors. This scalar
product will be especially effective in searching for heavy masses in
a jet clustering history when we evaluate it for combinations of
final-state objects.

The input to the Lorentz Layer (LoLa) network~\cite{Butter:2017cot} are
sets of 4-vectors $k_{\mu,i}$ with $i=1,...,N$. As a first step we
apply a combination layer to define linear combinations of the
4-vectors,
\begin{align}
k_{\mu,i} \stackrel{\text{CoLa}}{\longrightarrow}
\widetilde{k}_{\mu,j} 
= k_{\mu,i} \; C_{ij} 
\quad \text{with} \quad
C = 
 \begin{pmatrix}
  1 & \cdots & 0      & C_{1,N+1} & \cdots & C_{1,M} \\[-2mm]
  \vdots & \ddots & \vdots & \vdots &  & \vdots \\
  0 & \cdots & 1 &C_{N,N+1} & \cdots & C_{N,M} 
 \end{pmatrix} \; .
\end{align}
Here we use $N=60$ and $M=90$.
In a second step we transform all 4-vectors into measurement-motivated
objects,
\begin{align}
\tilde{k}_j 
\stackrel{\text{LoLa}}{\longrightarrow}
\hat{k}_j = 
 \begin{pmatrix*}[r]
 	m^2(\tilde k_j)\\ 
  p_T(\tilde k_j)\\ 
  w^{(E)}_{jm} \,E(\tilde k_m)\\ 
  w^{(p_T)}_{jm} \,p_T(\tilde k_m)\\ 
  w^{(m^2)}_{jm} \,m^2(\tilde k_m)\\ 
  w^{(d)}_{jm} \, d^2_{jm}\\ 
 \end{pmatrix*} \; ,
\label{eq:lola}
\end{align}
where $d^2_{jm}$ is the Minkowski distance between two 4-momenta
$\tilde k_j$ and $\tilde k_m$, and we either sum or minimize over the
internal indices.  One copy of the sum and five copies of the minimum
term are used.  Just for amusement we have checked with what precision
the network can learn the Minkowski metric from top vs QCD jet
data\cite{Butter:2017cot}. The LoLa network setup is then
straightforward, three fully connected hidden layers with 100, 50, and
2 nodes, and using the Adam optimizer~\cite{Kingma:2014vow}

\subsubsection{Latent Dirichlet Allocation}

Latent Dirichlet Allocation (LDA) is a widely used unsupervised
learning technique used in generative modelling for collections of
text documents~\cite{Blei:2003:LDA:944919.944937}. It can also uncover
the latent thematic structures in jets or events by searching for
co-occurrence patterns in high-level substructure
features~\cite{Dillon:2019cqt}.  Once the training is performed the
result is a set of learned themes, \ie probability distributions over
the substructure observable space. For the top tagger, we use a
two-theme LDA model, aimed at separate themes describing the signal
and background. We then use these distributions to infer theme
proportions and tag jets.  Both the training and inference are
performed with the Gensim software~\cite{rehurek_lrec}.

Before training, we pre-process the jets and map them to a
representation suitable for the Gensim software.  After clustering the
jets with the Cambridge/Aachen algorithm with a large cone radius,
each jet is iteratively unclustered without discarding any of the
branches in the unclustering history.  At each step of unclustering we
compute a set of substructure observables from the parent subjet and
the two daughter subjets, namely the subjet mass, the mass drop, the
mass ratio of the daughters, the angular separation of the daughters,
and the helicity angle of the parent subjet in the rest frame of the
heaviest daughter subjet. These five quantities are collated into a
5-dimensional feature vector for each node of the tree until the
unclustering is done.  This represents each jet as a list of 5-vector
substructure features.  To improve the top tagging we also include
complementary information, namely the $N$-subjettiness observables,
$(\tau_3/\tau_2,\tau_3/\tau_1,\tau_2/\tau_1)$. All eight observables
are binned and mapped to a vocabulary that is used to transform the
jet into a training document for the two-theme LDA.

LDA is generally used as an unsupervised learning technique. For this
study we use, for the sake of comparison, the LDA algorithm in a
supervised learning mode.  Regardless of whether the training is
performed in a supervised or unsupervised manner, once it is complete
it is straightforward to study what has been learned by the LDA
algorithm by inspecting the theme distributions.  For instance, in
Ref.~\cite{Dillon:2019cqt} the uncovered latent themes from the data
are plotted and one can easily distinguish QCD-like features in one
theme and top-like features in the other.

\subsubsection{Energy Flow Polynomials}


Energy Flow Polynomials (EFPs)~\cite{Komiske:2017aww} are a collection
of observables designed to form a linear basis of all infrared- and
collinear- (IRC) safe observables, building upon a rich literature of
energy
correlators~\cite{Tkachov:1995kk,Larkoski:2013eya,Moult:2016cvt}.  The
EFPs naturally enable linear methods to be applied to collider physics
problems, where the simplicity and convexity of linear models is
highly desirable.

EFPs are energy correlators whose angular structures are in
correspondence with non-isomorphic multigraphs.  Specifically, they
are defined using the transverse momentum fractions, $z_i =
p_{T,i}/\sum_j p_{T,j}$, of the $M$ constituents as well as their
pairwise rapidity-azimuth distances, $\theta_{ij} = ((y_i-y_j)^2 +
(\phi_i - \phi_j)^2)^{\beta/2}$, without any additional preprocessing:
\begin{align}
\text{EFP}_G = \sum_{i_1=1}^M \cdots \sum_{i_N=1}^M z_{i_1} \cdots z_{i_N} \prod_{(k,\ell)\in G} \theta_{i_k i_\ell},
\end{align}
where $G$ is a given multigraph, $N$ is the number of vertices in $G$,
and $(k,\ell)$ is an edge connecting vertices $k$ and $\ell$ in $G$.
The EFP-multigraph correspondence yields simple visual rules for
translating multigraphs to observables: vertices contribute an energy
factor and edges contribute an angular factor.  Beyond this, the
multigraphs provide a natural organization of the EFP basis when
truncating in the number of edges (or angular factors) $d$, with
exactly 1000 EFPs with $d\le 7$.

For the top tagging EFP model, all $d\le 7$ EFPs computable in
$\mathcal O(M^3)$ or faster are used (995 observables total) with
angular exponent $\beta = 1/2$.  Linear classification was performed
with Fisher's Linear Discriminant~\cite{fisher1936use} using
Scikit-learn~\cite{scikit-learn}.  Implementations of the EFPs are
available in the EnergyFlow package~\cite{energy_flow}.

\subsubsection{Energy Flow Networks}

An Energy Flow Network (EFN)~\cite{Komiske:2018cqr} is an architecture
built around a general decomposition of IRC-safe observables that
manifestly respects the variable-length and permutation-invariance
symmetries of observables.  Encoding the proper symmetries of particle
collisions in an architecture results in a natural way of processing
the collider data.

Any IRC-safe observable can be approximated arbitrarily well as
\begin{align}\label{eq:efndecomp}
\text{EFN} = F\left(\sum_{i=1}^M z_i \Phi(y_i, \phi_i)\right)  \; ,
\end{align}
where $\Phi:\mathbb R^2\to\mathbb R^\ell$ is a per-constituent mapping
that embeds the locations of the constituents in a latent space of
dimension $\ell$.  Constituents are summed over in this latent space to
obtain an event representation, which is then mapped by $F$ to the
target space of interest.

\begin{figure}[b!]
\centering
\includegraphics[width=0.6\textwidth]{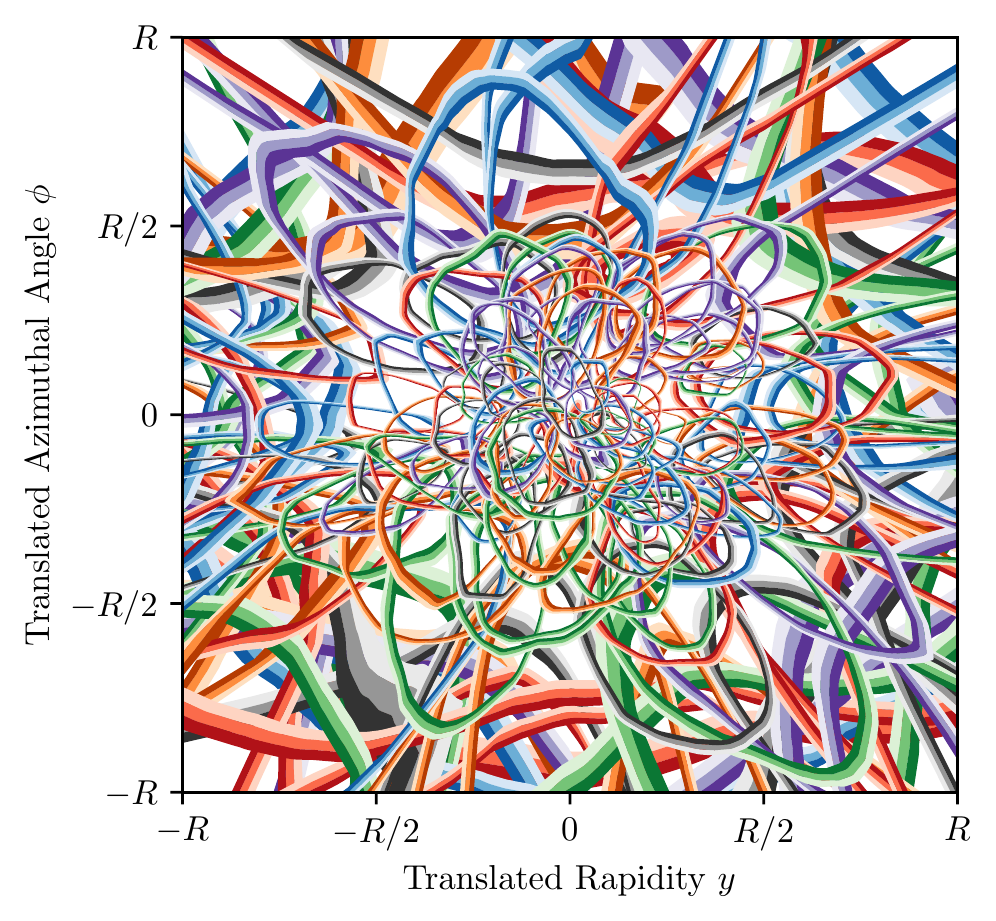}
\caption{Visualization of the trained top tagging EFN. Each contour
  corresponds to a filter, which represents the learned local latent
  space. The smaller filters probe the core of the jet and larger
  filters the periphery.  Figure from Ref.~\cite{Komiske:2018cqr}.}
\label{fig:efnfig}
\end{figure}

The EFN architecture parametrizes the functions $\Phi$ and $F$ with
neural networks, with specific implementation details of the top
tagging EFN architecture given in the Particle Flow Network section.
For both the top tagging EFNs and PFNs, input constituents are translated
to the origin of the rapidity-azimuth plane according to the
$p_T$-weighted centroid, rotated in this plane to consistently align
the principal axis of the energy flow, and reflected to locate the
highest $p_T$ quadrant in a consistent location.  A fascinating aspect
of the decomposition in Eq.~\eqref{eq:efndecomp} is that the learned
latent space can be examined both quantitatively and visually, as the
rapidity-azimuth plane is two dimensional and can be viewed as an
image.  Figure~\ref{fig:efnfig} shows a visualization of the top
tagging EFN, which learns a dynamic pixelization of the space.

\subsubsection{Particle Flow Networks}

Particle Flow Networks (PFNs)~\cite{Komiske:2018cqr} generalize the
EFNs beyond IRC safety.  In doing so, they make direct contact with
machine learning models on learning from point clouds, in particular
the Deep Sets framework~\cite{DBLP:conf/nips/ZaheerKRPSS17}.  This
identification of point clouds as the machine learning data structure
with intrinsic properties most similar to collider data provides a new
avenue of exploration.

In the collider physics language, the key idea is that any observable
can be approximated arbitrarily well as
\begin{align}
\text{PFN} = F\left(\sum_{i=1}^M \Phi(p_i)\right) \; ,
\end{align}
where $p_i$ contains per-particle information, such as momentum,
charge, or particle type.  Similar to the EFN case, $\Phi$ maps from
the particle feature space into a latent space and $F$ maps from the
latent space to the target space.  The per-particle features provided
to the network can be varied to study their information content. Only
constituent 4-momentum information is used here. Exploring the
importance of particle-type information for top tagging is an
interesting direction for future studies.

The PFN architecture parameterizes the functions $\Phi$ and $F$ with
neural networks.  For the top tagging EFN and PFN models, $\Phi$ and
$F$ are parameterized with simple three-layer neural networks of
$(100,100,256)$ and $(100,100,100)$ nodes in each layer, respectively,
corresponding to a latent space dimension of $\ell = 256$.  A ReLU
activation is used for each dense layer along with He-uniform
parameter initialization~\cite{heuniform}, and a two-node classifier
output is used with SoftMax activation, trained with
Keras~\cite{keras} and TensorFlow~\cite{tensorflow}.  Implementations
of the EFNs and PFNs are available in the EnergyFlow
package~\cite{energy_flow}.

\section{Comparison}
\label{sec:results}

\begin{figure}[t]
\centering
\includegraphics[width=0.7\textwidth]{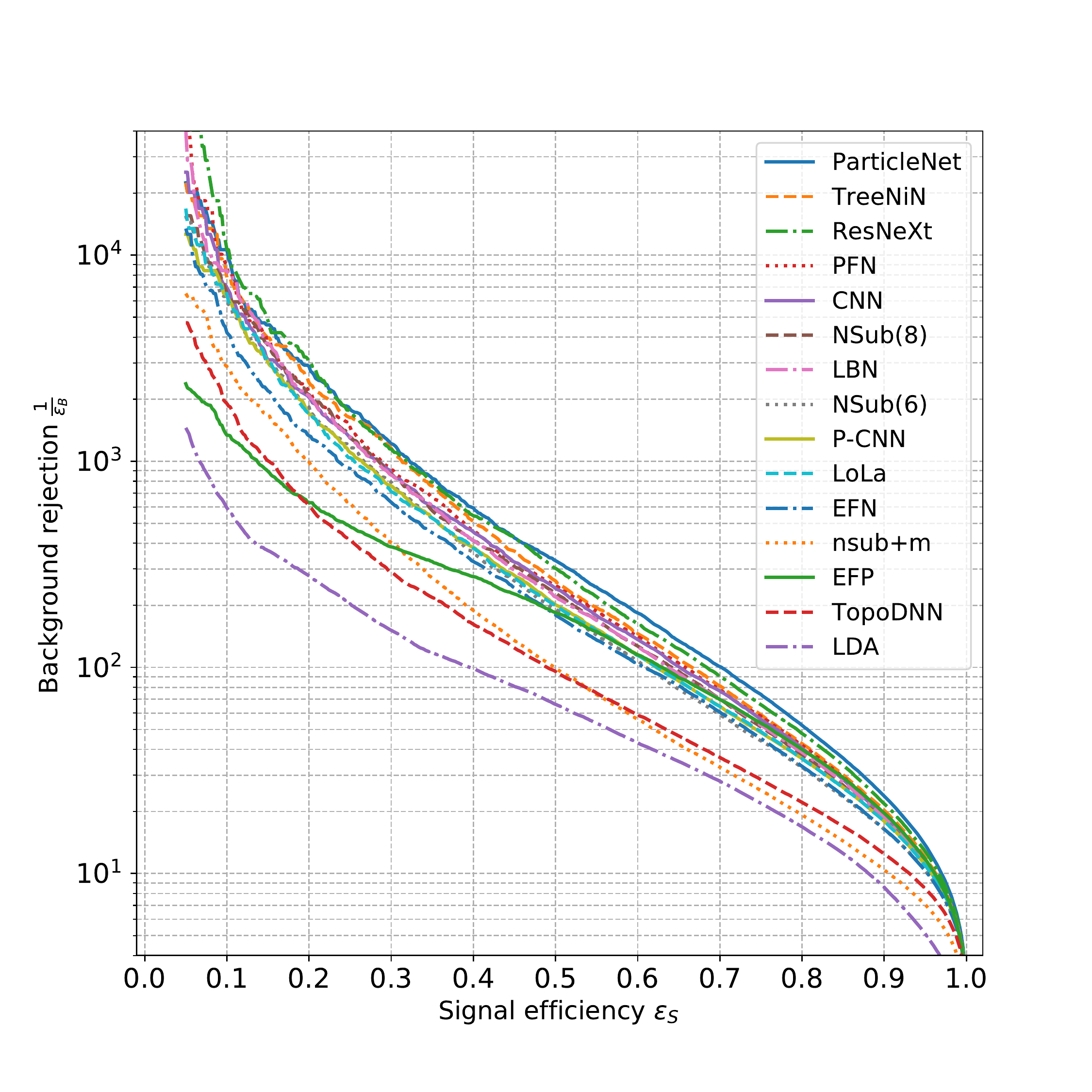}
\caption{ROC curves for all algorithms evaluated on the same test
  sample, shown as the AUC ensemble median of multiple trainings.
  More precise numbers as well as uncertainty bands given by the
  ensemble analysis are given in Tab.~\ref{tab:overview}.}
\label{fig:rocs}
\end{figure}

To assess the performance of different algorithms we first look at the
individual ROC curves over the full range of top jet signal
efficiencies. They are shown in Fig.~\ref{fig:rocs}, compared to a
simple tagger based on $N$-subjettiness~\cite{Thaler:2010tr} and jet
mass. We see how, with few exceptions, the different
taggers define similar shapes in the signal efficiency vs background
rejection plane.

Given that observation we can instead analyze three single-number
performance metrics for classification tasks. First, we compute the
\textsl{area under the ROC curve} shown in Fig.~\ref{fig:rocs}.  It is
bounded to be between 0 and 1, and stronger classification corresponds
to values larger than 0.5 at a chosen working point.  Next, the
\textsl{accuracy} is defined as the fraction of correctly classified
jets.  Finally, for a typical analysis application the
\textsl{rejection power} at a realistic working point is most
relevant. We choose the background rejection at a signal efficiency of
30\%.

All three figures of merit are shown in Tab.~\ref{tab:overview}. Most
approaches achieve an AUC of approximately 0.98 with the strongest
performance from the 4-vector-based ParticleNet, followed by the
image-based ResNeXt, the 4-vector-based TreeNiN, and the
theory-inspired Particle Flow Network. These approaches also reach the
highest accuracy and background rejection at fixed signal
efficiency. A typical accuracy is 93\%, and the quoted differences
between the taggers are unlikely to define a clear experimental
preference.

Instead of extracting these performance measures from single models we
can use ensembles. For this purpose we train nine models for each
tagger and define 84 ensemble taggers, each time combining six of
them. They allow us to evaluate the spread of the ensemble taggers
and define mean-of-ensemble and median-of-ensemble results.  We find
that ensembles leads to a 5~...~15\% improvement in performance,
depending on the algorithm. For the uncertainty estimate of the
background rejection we remove the outliers. In
Tab.~\ref{tab:overview} we see that the background rejection varies
from around 1/600 to better than 1/1000. For the ensemble tagger the
ParticleNet, ResNeXt, TreeNiN, and PFN approaches again lead to the
best results. Phrased in terms of the improvement in the
signal-to-background ratio they give factors $\epsilon_S/\epsilon_B >
300$, vastly exceeding the current top tagging performance in ATLAS
and CMS.

Altogether, in Fig.~\ref{fig:rocs} and Tab.~\ref{tab:overview} we see
that some of the physics-motivated setups remain competitive with the
technically much more advanced ResNeXt and ParticleNet
networks. This suggests that even for a straightforward task like top
tagging in fat jets we can develop efficient physics-specific
tools. While their performance does not quite match the
state-of-the-art standard networks, it is close enough to test both
approaches on key requirements in particle physics, like treatment of
uncertainties, stability with respect to detector effects,
etc.\medskip

\begin{table}[t]
\setlength{\tabcolsep}{5pt}
\centering
\begin{small}
\begin{tabular}{l|l|l|rrr|r}
\toprule
& AUC   & Acc  & \multicolumn{3}{c|}{$1/\epsilon_B \; (\epsilon_S=0.3)$} & $\#$Param \\
&&& single & mean & median & \\
\midrule
CNN~\cite{Macaluso:2018tck}                       & 0.981 & 0.930 & 914$\pm$14  &  995$\pm$15 &  975$\pm$18 &  610k \\
ResNeXt\cite{DBLP:journals/corr/XieGDTH16}        & 0.984 & 0.936 & 1122$\pm$47 & 1270$\pm$28 & 1286$\pm$31 & 1.46M \\
\midrule
TopoDNN~\cite{Pearkes:2017hku}                    & 0.972 & 0.916 & 295$\pm$5   &  382$\pm$\phantom{0}5 & 378 $\pm$\phantom{0}8 & 59k \\
Multi-body $N$-subjettiness 6~\cite{Moore:2018lsr}& 0.979 & 0.922 & 792$\pm$18  &  798$\pm$12 &  808$\pm$13 & 57k   \\
Multi-body $N$-subjettiness 8~\cite{Moore:2018lsr}& 0.981 & 0.929 & 867$\pm$15  &  918$\pm$20 &  926$\pm$18 & 58k   \\
TreeNiN~\cite{TreeNiNpaper}                       & 0.982 & 0.933 & 1025$\pm$11 & 1202$\pm$23 & 1188$\pm$24 & 34k  \\
P-CNN                                             & 0.980 & 0.930 & 732$\pm$24 &   845$\pm$13 &  834$\pm$14 & 348k  \\
ParticleNet~\cite{Qu:2019gqs}                     & 0.985 & 0.938 & 1298$\pm$46 & 1412$\pm$45 & 1393$\pm$41 & 498k  \\
\midrule
LBN~\cite{Erdmann:2018shi}                        & 0.981 & 0.931 & 836$\pm$17  &  859$\pm$67   &  966$\pm$20 & 705k   \\
LoLa~\cite{Butter:2017cot}                        & 0.980 & 0.929 & 722$\pm$17  &  768$\pm$11   &  765$\pm$11 & 127k   \\
LDA~\cite{Dillon:2019cqt}                         & 0.955 & 0.892 & 151$\pm$0.4 & 151.5$\pm$0.5 & 151.7$\pm$0.4 &  184k  \\
Energy Flow Polynomials~\cite{Komiske:2017aww}    & 0.980 & 0.932 &  384\phantom{$\pm$12} &     &   & 1k   \\
Energy Flow Network~\cite{Komiske:2018cqr}        & 0.979 & 0.927 & 633$\pm$31  &  729$\pm$13   &  726$\pm$11 & 82k   \\
Particle Flow Network~\cite{Komiske:2018cqr}      & 0.982 & 0.932 & 891$\pm$18  & 1063$\pm$21   & 1052$\pm$29 &  82k  \\
\midrule
GoaT & $0.985$ & $0.939$ & 1368$\pm$140 & & 1549$\pm$208 & 35k \\
\bottomrule
\end{tabular}
\end{small}
\caption{Single-number performance metrics for all algorithms
  evaluated on the test sample. We quote the area under the ROC curve
  (AUC), the accuracy, and the background rejection at a signal
  efficiency of $30\%$.  For the background rejection we also show the
  mean and median from an ensemble tagger setup. The number of
  trainable parameters of the model is given as well.  Performance
  metrics for the GoaT meta-tagger are based on a subset of events.}
\label{tab:overview}
\end{table}

\begin{figure}[t]
  \includegraphics[width=\textwidth]{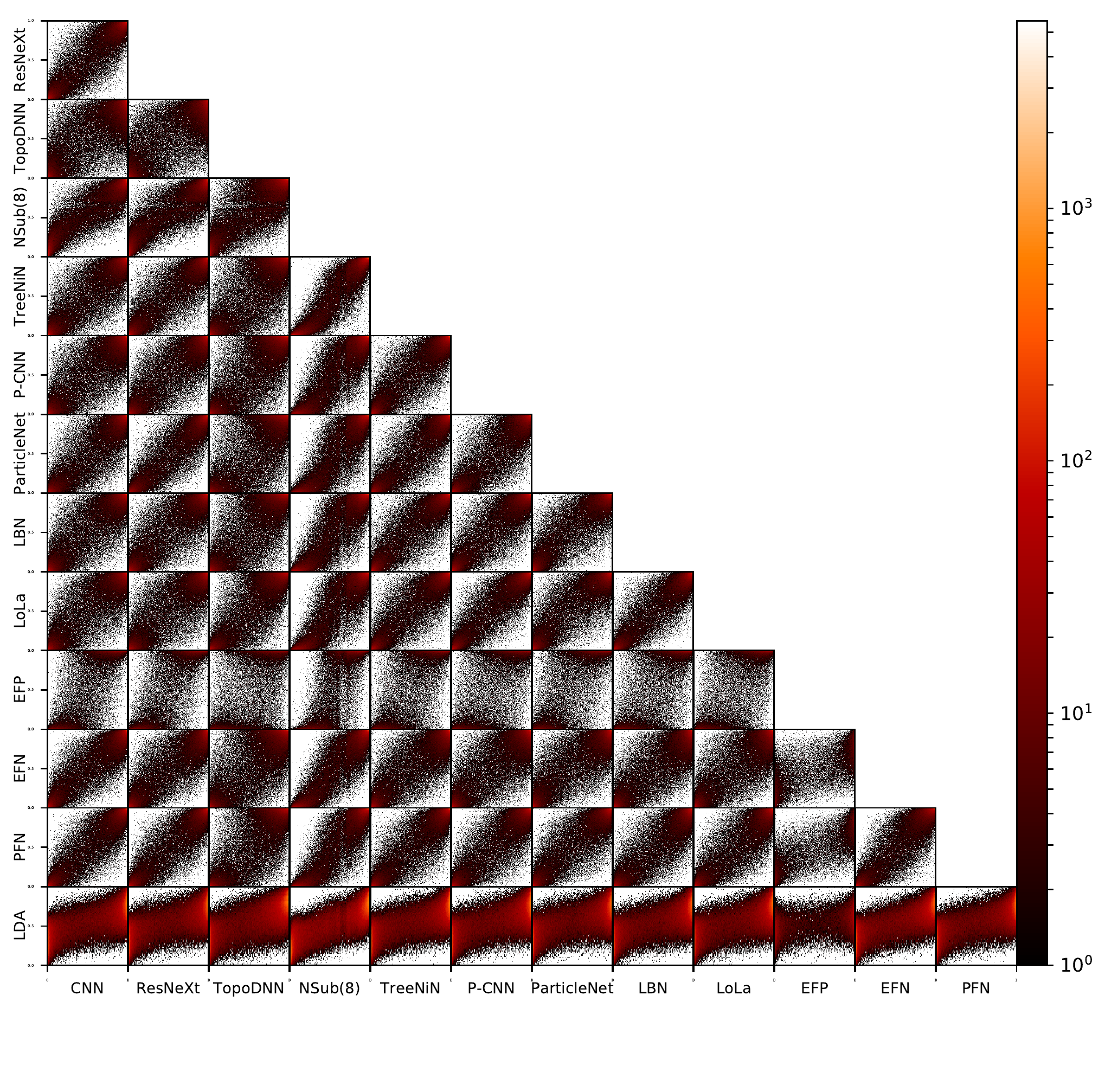}
  \caption{Pairwise distributions of classifier outputs, each in the
    range $0~...~1$ from pure QCD to pure top; the lower left corners
    include correctly identified QCD jets, while the upper right
    corners are correctly identified top jets. 
    LDA outputs are rescaled by a factor of $\approx 4$ to have the same range as other classifiers.}
\label{fig:corr}
\end{figure}

\begin{table}[t]
\setlength{\tabcolsep}{6pt}
\centering
\begin{footnotesize}
\begin{tabular}{l|rrrrrrrrrrrrrr}
\toprule
&   \multicolumn{2}{l}{CNN} &   \multicolumn{2}{l}{TopoDNN} &   \multicolumn{2}{l}{TreeNiN} &   \multicolumn{2}{l}{ParticleNet} &   \multicolumn{2}{l}{LoLa} &   \multicolumn{2}{l}{EFN} &   \multicolumn{2}{l}{LDA} \\
&    \multicolumn{2}{r}{$\big|$ \hfill ResNeXt} & \multicolumn{2}{r}{$\big|$ \hfill NSub(8)} & \multicolumn{2}{r}{$\big|$ \hfill P-CNN} & \multicolumn{2}{r}{$\big|$ \hfill LBN} & \multicolumn{2}{r}{$\big|$ \hfill EFP} & \multicolumn{2}{r}{$\big|$ \hfill PFN} &
\multicolumn{2}{l}{$\big|$} \\
\midrule
\hline
 CNN                               & 1    &     .983 &     .955 &     .978 &     .977 &   .973 &         .979 & .976 &  .974 & .973 & .981 & .984 & .784 \\
 ResNeXt                           & .983 &     1     &     .953 &     .977 &     .98  &   .975 &         .985 & .977 &  .974 & .975 & .976 & .983 & .782 \\
 TopoDNN                           & .955 &     .953 &     1     &     .962 &     .958 &   .962 &         .953 & .961 &  .97  & .945 & .955 & .961 & .777 \\
 NSub(8)                           & .978 &     .977 &     .962 &     1     &     .982 &   .975 &         .975 & .977 &  .977 & .964 & .979 & .977 & .802 \\
 TreeNiN                           & .977 &     .98  &     .958 &     .982 &     1     &   .98  &         .982 & .982 &  .981 & .968 & .973 & .981 & .786 \\
 P-CNN                             & .973 &     .975 &     .962 &     .975 &     .98  &   1     &         .978 & .98  &  .984 & .964 & .968 & .98  & .781 \\
 ParticleNet                       & .979 &     .985 &     .953 &     .975 &     .982 &   .978 &         1     & .978 &  .977 & .97  & .972 & .981 & .778 \\
 LBN                               & .976 &     .977 &     .961 &     .977 &     .982 &   .98  &         .978 & 1     &  .984 & .968 & .972 & .98  & .784 \\
 LoLa                              & .974 &     .974 &     .97  &     .977 &     .981 &   .984 &         .977 & .984 &  1     & .968 & .971 & .981 & .782 \\
 EFP                               & .973 &     .975 &     .945 &     .964 &     .968 &   .964 &         .97  & .968 &  .968 & 1     & .968 & .977 & .764 \\
 EFN                               & .981 &     .976 &     .955 &     .979 &     .973 &   .968 &         .972 & .972 &  .971 & .968 & 1     & .981 & .792 \\
 PFN                               & .984 &     .983 &     .961 &     .977 &     .981 &   .98  &         .981 & .98  &  .981 & .977 & .981 & 1     & .784 \\
 LDA                               & .784 &     .782 &     .777 &     .802 &     .786 &   .781 &         .778 & .784 &  .782 & .764 & .792 & .784 & 1     \\
\bottomrule
\end{tabular}
\end{footnotesize}
\caption{Correlation coefficients from the combined GoaT analyses}
\label{tab:corr}
\end{table}

The obvious question in any deep-learning analysis is if the tagger
captures all relevant information. At this point we have checked that
including full or partial information on the event-level kinematics of
the fat jets in the event sample has no visible impact on our quoted
performance metrics.  We can then test how correlated the classifier
output of the different taggers are, leading to the pair-wise
correlations for a subset of classifier outputs shown in
Fig.~\ref{fig:corr}. The correlation matrix is given in
Tab.~\ref{tab:corr}. As expected from strong classifier performances,
most jets are clustered in the bottom left and top right corners,
corresponding to identification as background and signal,
respectively. The largest spread is observed for correlations with the
EFP. Even the two strongest individual classifier outputs with
relatively little physics input --- ResNeXt and ParticleNet --- are
not perfectly correlated.

Given this limited correlation, we investigate whether a meta-tagger
might improve performance. Note that this GoaT (Greatest of all
Taggers) meta-tagger should not be viewed as a potential analysis
tool, but rather as a benchmark of how much unused information is
still available in correlations.  It is implemented as a fully
connected network with 5 layers containing 100-100-100-20-2 nodes. All
activation functions are ReLu, apart from the final layer's
SoftMax. Training is performed with the Adam~\cite{Kingma:2014vow}
optimizer with an initial learning rate of 0.001 and binary
cross-entropy loss.  We train for up to 50 epochs, but terminate if
there is no improvement in the validation loss for two consecutive
epochs, so a typical training ends after 5 epochs. The training data
is provided by individual tagger output 
 on the previous test sample
and split intro three subsets: GoaT-training (160k events),
GoaT-testing (160k events) and GoaT-validation (80k events).  We
repeat training/testing nine times, re-shuffling the events randomly
between the three subsets for each repetition. The standard deviation
of these nine repetitions is reported as uncertainty for GoaT taggers
in Tab.~\ref{tab:overview}. We show two GoaT versions, one using a
single output value per tagger as input (15 inputs), and one using all
values per tagger as input (135 inputs). 
All described taggers are used as input except LDA as it did not improve performance.
We see that the GoaT
combination improves the best individual tagger by more than $10\%$ in
background rejection, providing us with a realistic estimate of the
kind of improvement we can still expect for deep-learning top
taggers.\medskip

In spite of the fact that our study gives some definite answers
concerning deep learning for simple jet classification at the LHC, a
few questions remain open: first, we use jets in a relatively narrow
and specific $p_T$-slice. Future efforts could explore softer jets,
where the decay products are not necessarily inside one fat jet; higher
$p_T$, where detector resolution effects become crucial; and wider
$p_T$ windows, where stability of taggers becomes relevant. The
samples also use a simple detector simulation and do not contain
effects from underlying event and pile-up.

Second, our analysis essentially only includes calorimeter information
as input. Additional information exists in the distribution of tracks
of charged particles and especially the displaced secondary vertices
from decays of $B$-hadrons. How easily this information can be included
might well depend on details of the network architecture.

Third, when training machine learning classifiers on simulated events
and evaluating them on data, there exists a number of systematic
uncertainties that need to be considered. Typical examples are jet
calibration, MC generator modeling, and IR-safety when using theory
predictions. Understanding these issues will be a crucial next
step. Possibilities are to include uncertainties in the training or to
train on data in a weakly supervised or unsupervised fashion~\cite{weakly}.

Finally, we neglect questions such as computational complexity,
evaluation time and memory footprint. These will be important
considerations, especially once we want to include deep networks in
the trigger. A related questions will be how many events we need to
saturate the performance of a given algorithm.

\section{Conclusion}
\label{sec:other}

Because it is experimentally and theoretically well defined, top
tagging is a prime benchmark candidate to determine what we can expect
from modern machine learning methods in classification tasks relevant
for the LHC.  We have shown how different neural network architectures
and different representations of the data distinguish hadronically
decaying top quarks from a background of light quark or gluon jets.

We have compared three different classes of deep learning taggers:
image-based networks, 4-vector-based networks, and taggers relying on
additional considerations from relativistic kinematics or theory
expectations. We find that each of these approaches provide
competitive taggers with comparable performance, making it clear that
there is no golden deep network architecture.  Simple tagging
performance will not allow us to identify the kind of network
architectures we want to use for jet classification tasks at the
LHC. Instead, we need to investigate open questions like versatility
and stability in the specific LHC environment with the specific
requirements of the particle physics community for example related to
calibration and uncertainty estimates.

This result is a crucial step in establishing deep learning at
the LHC. Clearly, jet classification using deep networks working on
low-level observables is the logical next step in subjet analysis
to be fully exploited by
ATLAS and CMS. On the other hand, there exist a
range of open questions related to training, calibration, and
uncertainties.  They are specific to LHC physics and might well
require particle physics to move beyond applying standard tagging
approaches and in return contribute to the field of deep
learning. Given that we have now understood that different network
setups can achieve very similar performance for mostly calorimeter
information, it is time to tackle these open questions.

\bigskip
\begin{center} \textbf{Acknowledgments} \end{center} 

First and foremost we would like to thank Jesse Thaler for his
insights and his consistent and energizing support, and Michael
Russell for setting up the event samples for this comparison. We also
would like to thank Sebastian Macaluso and Simon Leiss for preparing
the plots and tables in this paper.  In addition, we would like to
thank all organizers of the ML4Jets workshops in Berkeley and at FNAL
for triggering and for supporting this study. Finally, we are also
grateful to Michel Luchmann for providing us with
Fig.~\ref{fig:averaged_images}.  PTK and EMM were supported by the
Office of Nuclear Physics of the U.S. Department of Energy (DOE) under
grant DE-SC-0011090 and by the DOE Office of High Energy Physics under
grant DE-SC-0012567, with cloud computing resources provided through a
Microsoft Azure for Research Award.  The work of BN is supported by
the DOE under contract DE-AC02-05CH11231.  KC and SM are supported
from The Moore-Sloan Data Science Environment, and NSF OAC-1836650 and
NSF ACI-1450310 grants. SM gratefully acknowledges the support of
NVIDIA Corporation with the donation of a Titan V GPU used for this
project.


\end{document}